\documentclass[11pt]{article}
\usepackage[english,activeacute]{babel}
\usepackage{epsfig}
\usepackage{amsmath}
\usepackage{epic}
\usepackage{amssymb}
\usepackage{fancybox}
\usepackage{wrapfig}
\usepackage[font=footnotesize,labelsep=period]{caption}
\usepackage{amsthm}
\usepackage{float}
\usepackage{xcolor}
\usepackage{booktabs}

\usepackage{wrapfig}

\usepackage[all]{xy}
\usepackage{xypic, color, graphics}

\theoremstyle{plain}

\theoremstyle{definition}

\newcommand{\beq}{\begin{eqnarray}}
\newcommand{\eeq}{\end{eqnarray}}
\newcommand{\ba}{\begin{array}}
\newcommand{\ea}{\end{array}}
\newcommand{\be}{\begin {equation}}
\newcommand{\ee}{\end{equation}}
\def\picture #1 by #2 (#3){
  \vbox to #2{
    \hrule width #1 height 0pt depth 0pt
    \vfill
    \special{picture #3} 
    }
  }
\def\scaledpicture #1 by #2 (#3 scaled #4){{
  \dimen0=#1 \dimen1=#2
  \divide\dimen0 by 1000 \multiply\dimen0 by #4
  \divide\dimen1 by 1000 \multiply\dimen1 by #4
  \picture \dimen0 by \dimen1 (#3 scaled #4)}
  }
  
\usepackage{fancyhdr}

\begin{document}

\centerline{\Large {\bf Iterative symmetry search and }} \vskip 0.20cm
\centerline{\Large {\bf reduction of a water wave model in $2+1$ dimensions }} \vskip 0.75cm

\centerline{ P.G. Est\'evez$^{\dagger}$, J.D.
Lejarreta$^{\ddagger}$ and C. Sard\'on$^{\dagger}$} \vskip 0.25cm

\centerline{$^{\dagger}$Department of Fundamental Physics, University of Salamanca,}
\medskip
\centerline{Plza. de la Merced s/n, 37.008, Salamanca, Spain.}
\medskip
\centerline{$^{\ddagger}$Department of Applied Physics, University of Salamanca,}
\medskip
\centerline{Plza. de la Merced s/n, 37.008, Salamanca, Spain.}
\medskip



\begin{abstract}
We present the iterative classical point symmetry analysis of a shallow water wave equation in $2+1$ dimensions and that of its corresponding nonisospectral, two component Lax pair. A few reductions arise and are identified  with celebrate equations in the Physics and Mathematics literature of nonlinear waves.
We pay particular attention to the isospectral or nonisospectral nature of the reduced spectral problems.
\end{abstract}



\section{Introduction}

Invariance of a differential equation under a group of transformations is synonymous of existence of {symmetry} and, consequently, of {conserved quantities} \cite{Nucci3}.
Such invariance helps us to achieve partial or complete integration of the equation.

A conserved quantity for a first-order differential equation can lead to its integration by quadrature, whilst
for higher-order ones, it leads to a reduction of their order \cite{Olver,Stephani}. 
Many of the existing solutions to physical phenomena described by differential equations have been obtained through symmetry arguments \cite{Olver,PS,Stephani}.
Nevertheless, finding conserved quantities is a nontrivial task.

The most famous and established method for finding point symmetries is the {classical Lie symmetry method} (CLS) developed by Lie in 1881 \cite{Lie81,Lie90,Lie93,Olver,Stephani}.
Although the CLS represents a very powerful tool, it yields cumbersome calculations to be solved by hand. Notwithstanding, the increased number of available software packages for symbolic calculus
has made of generalizations of the CLS analysis very tractable approaches to find conservation laws, explicit solutions, etc.

\noindent
If we impose that symmetries leave certain submanifold invariant, we find the class of {\it conditional symmetries} or the so-called {\it nonclassical symmetry method} (NSM) introduced in 1969 by
Bluman and Cole \cite{BC1} and later applied by many authors \cite{ArriBeck,ClarkWinter,EstLejaSar, Nucci1, NucciAmes}.
A remarkable difference between the CLS and NSM is that the latter provides us with 
no longer linear systems of differential equations from which to obtain the symmetries.

In this paper, we aim to perform the CLS on the so-called  {\it Bogoyavlenskii-Kadomtsev-Petviashvili equation} (denoted as $(2+1)$-BKP, henceforth) 
which takes the form \cite{EH00,Est1}
\begin{equation}\label{BKP}
\left(u_{xt}+u_{xxxy}+8u_xu_{xy}+4u_{xx}u_y\right)_x+\sigma u_{yyy}=0,\quad \sigma=\pm 1.
\end{equation}
This equation is a model for evolutionary shallow water waves represented by the scalar field $u(x,y,t)$ as the height of the wave.
In \cite{YTF88}, it was proved that this equation is a reduction of the well-known $(3+1)$-{\it Kadomtsev--Petviashvili equation} for the description
of wave motion \cite{Dru,KP}. Notice that the $(2+1)$-BKP is also a modified version of the {\it Calogero-Bogoyanlevski-Schiff equation} \cite{Bogo3,Calo,SE13}.

Equation \eqref{BKP} is integrable according to the Painleve property (PP) and admits a Lax (LP) \cite{EH00}.
Results on solutions of solitonic nature for \eqref{BKP} were pursued through the {\it singular manifold method} \cite{EH00,EP052,Weiss00,WTC}.
In particular, {\it lump solitons}, or solutions which decay polinomially in all directions, were found. The interest
of lump solutions roots in their nontrivial dynamics and interactions \cite{EP08,Est1,MS96}.

In particular, we focus on the study of its corresponding Lax pair in $2+1$ dimensions \cite{Est1}.
It is a complex, two component linear problem \cite{EH00}  
\begin{align}\label{plst}
        &\psi_{xx}=-i\psi_y-2u_x\psi,\nonumber\\
       &\psi_t=2i\psi_{yy}-4u_y\psi_x+(2u_{xy}+2i\omega_y)\psi
       \end{align}
\noindent
and its complex conjugate

\begin{align}\label{plstcc}
        &\chi_{xx}=i\chi_y-2u_x\chi,\nonumber\\
       &\chi_t=-2i\chi_{yy}-4u_y\chi_x+(2u_{xy}-2i\omega_y)\chi
       \end{align}

\noindent
with $\psi(x,y,t)$ and $\chi(x,y,t)$ being the eigenfunctions. 
The compatibility condition of the cross derivatives ($\psi_{xxt}=\psi_{txx}$) in \eqref{plst} retrives \eqref{BKP}.
So does the compatibility condition for ($\chi_{xxt}=\chi_{txx}$) in \eqref{plstcc}.

Notice that the spectral parameter is not present in the Lax pair. This does not necessarily imply that it is an isospectral Lax pair.
Indeed, there exists a gauge transformation
which allows us to express the nonisospectral spectral linear problem as a spectral parameter-free linear problem. The converse is possible
by introducing ``$\lambda$'' or  {\it spectral parameter}, conveniently.

The importance of reduced spectral problems and reduced spectral parameters resides 
in the fact that in $1+1$ dimensions, it is not usual to find nonisospectral versions. The nonisospectrality of $1+1$ dimensional Lax pairs gives rise to some inconveniency, since 
the {\it inverse scattering transform} (IST) \cite{AC91,AS81} can no longer be worked upon them, for example.
If the IST cannot be used, the possibilities of solving the nonlinear equation through an associated spectral problem, are diminished.
On the other hand, many of the 
Lax pairs found in $2+1$ dimensions are nonisospectral \cite{EstLejaSar, EP052}.
 For this matter, we pay closer attention at the reductions of the associated spectral problem.
In this way, we stress out the importance of finding isospectral Lax pairs in $1+1$ dimensions and the surprising nature of nonisospectral ones.

In section 2, we introduce the CLS for partial differential equations (PDEs)
and apply it to the $(2+1)$-BKP and its corresponding Lax pair.
In Section 3, we will obtain a classification of its possible reductions to $1+1$ dimensions of the equation and the Lax pair, depending on different values of the arbitrary functions
appearing in the obtained symmetries.
We will identify six interesting reductions. Two of them will be nontrivial. One of such nontrivial reduction corresponds with the celebrated {Korteweg de Vries equation} (KdV) \cite{Drazin1,KdV}. 
The second reduction found will be submitted to a second CLS calculation.
So, Section 4 shall be devoted to the classical symmetry computation of the aforementioned second nontrivial reduction in $1+1$ dimensions.
A list of four reductions to ODEs will be displayed, considering different values for the constants of integration appearing in the symmetry computation.
To conclude we shall enclose a summary of the most relevant results found throughout the ``iterative symmetry search and reduction" procedure.

\section{The nonclassical symmetry approach}


As a matter of convenience, we shall rewrite \eqref{BKP} in the conservative form
\begin{align}\label{cbkp}
        &u_{xt}+u_{xxxy}+8u_xu_{xy}+4u_{xx}u_y=\omega_{yy},\nonumber\\
       &u_y=\omega_x
\end{align}

\noindent
where we have introduced the auxiliary scalar field $\omega(x,y,t)$. 
We restrict ourselves to the case $\sigma=-1$. Results for $\sigma=1$ can be obtained by considering that if $u(x,y,t)$ is a solution for $\sigma=-1$, then $u(x,iy,it)$ is a solution for  $\sigma=1$.

A priori, we propose the most general Lie point symmetry transformation in which the coefficients of the infinitesimal generator
can depend on any dependent or independent variable

\begin{align}\label{bkplietransf}
x&\rightarrow x+\epsilon \xi_1(x,y,t,u,\omega)+O(\epsilon^2),\nonumber\\
y&\rightarrow y+\epsilon \xi_2(x,y,t,u,\omega)+O(\epsilon^2),\nonumber\\
t&\rightarrow t+\epsilon \xi_3(x,y,t,u,\omega)+O(\epsilon^2),\nonumber\\
u&\rightarrow u+\epsilon \eta_u(x,y,t,u,\omega)+O(\epsilon^2),\nonumber\\
\omega&\rightarrow \omega+\epsilon \eta_{\omega}(x,y,t,u,\omega)+O(\epsilon^2).
\end{align}
\noindent
If we search for general Lie point transformation of the Lax pair too, the spectral functions
must be transformed accordingly with

 \begin{align}\label{translax}
  \psi&\rightarrow\psi+\epsilon\eta_{\psi}(x,y,t,u,\omega,\psi,\chi)+O(\epsilon^2),\nonumber \\
  \chi&\rightarrow\chi+\epsilon\eta_{\chi}(x,y,t,u,\omega,\psi,\chi)+O(\epsilon^2).
 \end{align}

Associated with this transformation, there exists an infinitesimal generator
\begin{equation}\label{infgenbkp}
X=\xi_1\frac{\partial}{\partial x}+\xi_2\frac{\partial}{\partial y}+\xi_3\frac{\partial}{\partial t}+\eta_u\frac{\partial}{\partial u}+\eta_{\omega}\frac{\partial}{\partial \omega}+\eta_{\psi}\frac{\partial}{\partial \psi}+\eta_{\chi}\frac{\partial}{\partial \chi},
\end{equation}
where the subscripts in $\eta_p$ have been added according to the field ``$p$" to which each $\eta$ is associated.
This transformation must leave \eqref{plst}, \eqref{plstcc} and \eqref{cbkp}, invariant. From now, since equations \eqref{plst} and \eqref{plstcc} are equivalent, we shall exclude \eqref{plstcc}
from our calculus as a matter of simplification. Our results obtained for $\psi$ can be similarly extrapolated for $\chi$.

To proceed with CLS, we follow the next steps
\begin{enumerate}
\item Introduce the infinitesimal transformation \eqref{bkplietransf} and its further derivatives in equations \eqref{plst} and \eqref{cbkp}.
\item Select the linear terms in $\epsilon$ and set them equal to zero. The zero order in $\epsilon$ retrieves the original equations.
\item Substitute the values of $\psi_{xx},\psi_t,u_{xt},u_y$ from \eqref{plst} and \eqref{cbkp}, correspondingly.
Higher-order derivatives in these terms must be introduced according to these expressions.
\item From the steps above, we obtain an overdetermined system of differential equations for $\xi_1,\xi_2,\xi_3,\eta_{u},\eta_{\omega},\eta_{\psi}$ by
setting equal to zero terms in different orders of the derivatives of the fields, if we pursue Lie point symmetries.
\end{enumerate}

\noindent
Steps 1 to 5 are nontrivial, leading to cumbersome equations
which have been manipulated with a symbolic calculus package. In our case, we make use of MAPLE\textcopyright.

We extend transformation \eqref{bkplietransf} to first, second and higher-order derivatives appearing in the equations \eqref{cbkp}, 
\begin{align}
&u_x\rightarrow u_x+\epsilon \eta_u^{[x]}+O(\epsilon^2),\nonumber\\
&u_y\rightarrow u_y+\epsilon \eta_u^{[y]}+O(\epsilon^2),\nonumber\\
&\omega_x\rightarrow \omega_x+\epsilon \eta_{\omega}^{[x]}+O(\epsilon^2),\nonumber\\
&\omega_y\rightarrow \omega_y+\epsilon \eta_{\omega}^{[y]}+O(\epsilon^2),\nonumber\\
&u_{xx}\rightarrow  u_{xx}+\epsilon \eta_u^{[xx]}+O(\epsilon^2),\nonumber\\ 
&u_{xy}\rightarrow u_{xy}+\epsilon \eta_u^{[xy]}+O(\epsilon^2),\nonumber\\
&u_{xt}\rightarrow u_{xt}+\epsilon \eta_u^{[xt]}+O(\epsilon^2),\nonumber\\
&\omega_{yy}\rightarrow \omega_{yy}+\epsilon \eta_\omega^{[yy]}+O(\epsilon^2),\nonumber\\
&u_{xxxy}\rightarrow u_{xxxy}+\epsilon \eta_u^{[xxxy]}+O(\epsilon^2).
\end{align}

\noindent
And derivatives appearing in the Lax pair \eqref{plst}

\begin{align}
&\psi_{x}\rightarrow\psi_{x}+\epsilon \eta_{\psi}^{[x]}+O(\epsilon^2),\nonumber\\
 & \psi_{y}\rightarrow\psi_{y}+\epsilon \eta_{\psi}^{[y]}+O(\epsilon^2),\nonumber\\
 &\psi_{t}\rightarrow\psi_{t}+\epsilon \eta_{\psi}^{[t]}+O(\epsilon^2),\nonumber\\
&\psi_{xx}\rightarrow\psi_{xx}+\epsilon \eta_{\psi}^{[xx]}+O(\epsilon^2),\nonumber\\
& \psi_{yy}\rightarrow \psi_{yy}+\epsilon \eta_{\psi}^{[yy]}+O(\epsilon^2).
\end{align}

\noindent
The prolongations needed for the equation \eqref{cbkp} are $\eta_u^{[xxxy]}$, $\eta_u^{[xt]}$, $\eta_u^{[xy]}$, $\eta_u^{[xx]}$, $\eta_\omega^{[yy]}$, $\eta_u^{[x]}$, $\eta_u^{[y]}$, $\eta_\omega^{[x]}$, $\eta_\omega^{[y]}$
and for the Lax pair \eqref{plst}, the prolongations are $\eta_{\psi}^{[yy]}$, $\eta_{\psi}^{[xx]}$,
$\eta_{\psi}^{[t]}$, $\eta_{\psi}^{[y]}$, $\eta_{\psi}^{[x]}$. These prolongations can be calculated according to the Lie method explained in textbooks \cite{Stephani}.
\noindent
From the original equations, we make use of
\begin{align}\label{cbkp2}
&\omega_{yy}=u_{xt}+u_{xxxy}+8u_xu_{xy}+4u_{xx}u_y,\nonumber\\
&u_y=\omega_x,
\end{align}
\noindent
and the original the Lax pair
\begin{align}
        &\psi_{xx}=-i\psi_y-2u_x\psi,\nonumber \\
       &\psi_t=2i\psi_{yy}-4u_y\psi_x+(2u_{xy}+2i\omega_y)\psi.
       \end{align}
\noindent
Also, their further derivatives must be computed using the given expressions.

\noindent
Introducing such relations we arrive at the classical Lie symmetries 

\begin{align}\label{csymbkp}
&\xi_1=\frac{\dot A_3(t)}{4}x+A_1(t),\nonumber\\
&\xi_2=\frac{\dot A_3(t)}{2}y+A_2(t),\nonumber\\
&\xi_3=A_3(t),\nonumber\\
&\eta_u=-\frac{\dot A_3(t)}{4}u+\frac{\dot A_2(t)}{8}x+\frac{\ddot{A_3}(t)}{16}xy+\frac{\dot{A_1}(t)y}{4}+B_1(t),\nonumber\\
&\eta_{\omega}=-\frac{\dot A_3(t)}{2}w+\frac{\dot A_1(t)}{4}x+\frac{\ddot{A_3}(t)}{32}x^2+B_3(t)y+\frac{\ddot{A_2}(t)}{16}y^2+,\nonumber
\\
&\hspace{2em}+\frac{\dddot{A}_3(t)}{96} y^3+B_2(t),\nonumber\\
&\eta_{\psi}=\left[-2\lambda-\frac{\dot A_3(t)}{8}+i\left(\frac{\dot A_2(t)}{4}y+\frac{\ddot{A}_3(t)}{16}y^2+2\int{B_3(t)dt}\right)\right]\psi.
\end{align}

\noindent
These symmetries depend on a constant $\lambda$ and six arbitrary functions of time, $A_1(t),$ $A_2(t)$, $A_3(t)$ and $B_1(t),$ $B_2(t),$ $B_3(t)$, 
which shall serve us as a way to classify the possible reductions.
Indeed, we have the possible reductions attending to

\begin{table}[H]
\begin{center}
\begin{tabular}{lllclll}\toprule
\multicolumn{3}{c}{$\text{Case I}: A_3(t)\neq 0$}&&\multicolumn{3}{c}{$\text{Case II}: A_3(t)=0$}
\\ 
\midrule
1.&$A_1(t)\neq 0$& $A_2(t)\neq 0$&&1.& $A_1(t)\neq 0$&$A_2(t)\neq 0$\\
2.& $A_1(t)\neq 0$& $A_2(t)=0$&&2.& $A_1(t)\neq 0$&$A_2(t)=0$\\
3.& $A_1(t)=0$ & $A_2(t)\neq 0$&&3.& $A_1(t)=0$ & $A_2(t)\neq 0$\\
\bottomrule
\end{tabular}
 \end{center}
\caption{Reductions for $2+1$-BKP}
\label{Tab1}
\end{table}

\section{Reduction to $1+1$ dimensions}


To reduce the problem, we have to solve the Lagrange-Charpit system by the method of characteristics, that is integration of the 
{characteristic system} \cite{Olver,Stephani}
\begin{equation}\label{LCBKP21}
\frac{dx}{\xi_1}=\frac{dy}{\xi_2}=\frac{dt}{\xi_3}=\frac{du}{\eta_u}=\frac{d\omega}{\eta_{\omega}}=\frac{d\psi}{\eta_{\psi}}.
\end{equation}


We shall use the next notation for the reduced variables
\begin{equation}
 x,y,t\rightarrow x_1,x_2
\end{equation}
and for the reduced fields
\begin{align}
& \omega(x,y,t)\rightarrow \Omega(x_1,x_2),\nonumber\\
& u(x,y,t)\rightarrow U(x_1,x_2),\nonumber\\
& \psi(x,y,t)\rightarrow \Phi(x_1,x_2).
\end{align}
As a matter of simplification, we shall drop the dependency on $(x_1,x_2)$ of certain fields in the forthcoming expressions.

\begin{itemize}
\item {\it Case I.1. $A_3(t)\neq 0, A_1\neq 0, A_2\neq 0$}
\begin{itemize}
\item Reduced variables 
\begin{equation*}
x_1=\frac{x}{A_3(t)^{1/4}}-\int{\frac{A_1(t)}{A_3(t)^{5/4}}dt},\quad x_2=\frac{y}{A_3(t)^{1/2}}-\int{\frac{A_2(t)}{A_3(t)^{3/2}}dt}.
\end{equation*}
\item Reduced fields
\begin{align*}
u(x,y,t)&=\frac{U(x_1,x_2)}{A_3(t)^{1/4}}+\frac{x_1x_2}{16}\frac{\dot A_3(t)}{A_3(t)^{1/4}}
+\frac{x_1}{16}\left(2\frac{A_2(t)}{A_3(t)^{3/4}}+B(t)\frac{\dot A_3(t)}{A_3(t)^{1/4}}\right)+
\nonumber\\
&+\frac{x_2}{16}\left(4\frac{A_1(t)}{A_3(t)^{1/2}}+A(t)\frac{\dot A_3(t)}{A_3(t)^{1/4}}\right),
\end{align*}

\begin{align*}
\omega(x,y,t)&=\frac{\Omega(x_1,x_2)}{A_3(t)^{1/2}}+\frac{x_2^3}{192}\left(2A_3(t)^{1/2}\ddot{A_3}(t)-\frac{\dot{A}_3(t)^2}{A_3(t)^{1/2}}\right)+\frac{x_1^2}{32}\frac{\dot{A}_3(t)}{A_3(t)^{1/2}}+
\nonumber\\
&+\frac{x_1}{16}\left(4\frac{A_1(t)}{A_3(t)^{3/4}}+A(t)\frac{\dot{A}_3(t)}{A_3(t)^{1/2}}\right)+
\nonumber\\
&+\frac{x_2^2}{64}\left[4\dot{A_2}(t)-2\frac{A_2(t)\dot{A}_3(t)}{A_3(t)}+\frac{B(t)}{A_3(t)^{1/2}}\left(2A_3(t)\ddot{A_3}(t)-\dot{A}_3(t)^2\right)\right]+
\nonumber\\
&+ x_2\left[\frac{\int{B_3(t)dt}}{A_3(t)^{1/2}}-\frac{1}{16}\frac{A_2(t)^2}{A_3(t)^{3/2}}+\frac{B(t)}{16}\left(2\dot{A_2}(t)-\frac{A_2(t)\dot{A}_3(t)}{A_3(t)}\right)+
\right.\nonumber
\\
&\left.
+\frac{B(t)^2}{64}\left(2A_3(t)^{1/2}\ddot{A_3}(t)-\frac{\dot{A_3}(t)^2}{A_3(t)^{1/2}}\right)\right],
\end{align*}
where we have used the definitions
\begin{equation*}
A(t)=\int{\frac{A_1(t)}{A_3(t)^{5/4}}dt},\quad B(t)=\int{\frac{A_2(t)}{A_3(t)^{3/2}}dt}.
\end{equation*}
\item Reduced equation
\begin{align}
       &U_{x_2x_2}=\Omega_{x_1x_2},\nonumber\\
       &\Omega_{x_2x_2}=U_{x_2x_1x_1x_1}+8U_{x_2x_1}U_{x_1}+4U_{x_1x_1}U_{x_2}.
\end{align}
These two equations can be summarized in
\begin{equation}\label{rode}
U_{x_2x_1x_1x_1x_1}+4\Big((U_{x_2}U_{x_1})_{x_1}+U_{x_1}U_{x_1x_2}\Big)_{x_1}-U_{x_2x_2x_2}=0,
\end{equation}
\noindent
which appears in \cite{Wazwaz} and has {\it multiple soliton solutions}.

\item Reduced eigenfunction
\begin{align*}
\psi(x,y,t)&=\frac{\Phi(x_1,x_2)}{A_3(t)^{1/8}}\exp\left[i\frac{x_2^2}{16}\dot{A}_3(t)+i\frac{x_2}{8}\left(2\frac{A_2(t)}{A_3(t)^{1/2}}+B(t)\dot{A}_3(t)\right)-\right.\nonumber\\
%
&\hspace*{-2em}-2\lambda\int{\frac{dt}{A_3(t)}}\left.+i\int{\left(2\frac{\int{B_3(t)dt}}{A_3(t)}+\frac{B(t)}{4}\frac{\dot{A}_2(t)}{A_3(t)^{1/2}}+\frac{B(t)^2}{16}\ddot{A}_3(t)\right)dt}\right].
\end{align*}
\item Reduced Lax pair
\begin{align}\label{rodelp}
&\Phi_{x_1x_1}+i\Phi_{x_2}+2U_{x_1}\Phi=0,\nonumber\\
 &i\Phi_{x_2x_2}-2U_{x_2}\Phi_{x_1}+\left(U_{x_1x_2}+i\Omega_{x_2}+\lambda\right)\Phi=0.
 \end{align}
Here a constant of integration ``$\lambda$'' arises. This constant of integration plays the role of the spectral parameter
in $1+1$ dimensions. 

\end{itemize}

\item {\it The cases I.2. and I.3. will be omitted for being equivalent to I.1.}

\item{\it Case II.1. $A_3(t)=0$, $A_1(t)\neq 0$, $A_2(t)\neq 0$}

\begin{itemize}
\item Reduced variables
\begin{equation*}
x_1=\frac{A_2(t)x-A_1(t) y}{A_2(t)^{3/2}}-F(t)\quad x_2= \int \frac{A_1(t)}{A_2(t)^{5/2}}dt
\end{equation*}

\item Reduced fields
\begin{align*}
u(x,y,t)&=\frac{U(x_1,x_2)}{\sqrt{A_2(t)}} +\left(\frac{B_1(t)}{A_2(t)}+\frac{\dot A_2(t)}{8 A_2(t)}x \right) y +\\
+&\left(2\frac{\dot A_1(t)}{A_2(t)}-\frac{A_1(t) \dot A_2(t)}{A_2(t)^2}\right)\frac{y^2}{16} +\left(\frac{\int B_3(t)dt }{\sqrt{A_2(t)}}-\frac{A_1(t)^2}{8A_2(t)^{3/2}}\right)x_1
\end{align*}
\begin{align*}
w(x,y,t)&=-A_1(t)\frac{ \Omega(x_1,x_2)}{A_2(t)^{3/2}}+  \frac{\ddot A_2(t)}{48A_2(t)}y^3+\left(4\frac{B_3(t)}{A_2(t)}-\frac{A_1(t) \dot A_1(t)}{A_2(t)^2}\right)\frac{y^2}{8}+\\  +&\left(\frac{B_2(t)}{A_2(t)}+\frac{\dot A_1(t)}{4A_2(t)}x \right) y+\frac{\dot A_2(t)}{16}x_1^2+\\ 
&+\left (4i\lambda \frac{A_1(t)}{A_2(t)^{3/2}}+\frac{A_2(t) B_2(t)+4(\lambda - i \int B_3(t)dt )^2}{A_1(t)\sqrt{A_2(t)}}+  \frac{\dot A_1(t)A_2(t)}{4A_1(t)}F(t) \right)x_1
\end{align*}

with

\begin{align*}
&F(t)=\int\left(\frac{5}{2} \frac{A_1(t)^3}{A_2(t)^{7/2}}+ 4 \frac{B_1(t)}{A_2(t)^{3/2}}-12\frac{A_1(t)}{A_2(t)^{5/2}} \int B_3(t)dt \right)dt\\
&G(z_2)=U_{x_1}-\Omega_{x_1}
\end{align*}

\item Reduced equation
\begin{align*}
&U_{x_1x_1}-\Omega_{x_1x_1}=0\\
&U_{x_1x_1x_1x_1}+12U_{x_1x_1} U_{x_1}-U_{x_1x_2}=0.
\end{align*}
which can be equivalently rewritten as
\begin{equation}
 U_{x_1x_1x_1x_1x_1}+12 U_{x_1x_1}^2+ 12 U_{x_1}U_{x_1x_1x_1}-U_{x_1x_1x_2 }=0.
\end{equation}
This reduced equation corresponds with the KdV in $1+1$ dimensions \cite{Drazin}. 
{\it Therefore, we can conclude
that the $(2+1)$-BKP equation is a generalization of the KdV to $2+1$ dimensions.}

\item Reduced Eigenfunction
\begin{align*}
\Psi (x,y,t)&=\frac{\Phi (x_1,x_2)}{ A_2(t)^{1/4}}\exp\left( i\frac{\dot A_2(t)}{8A_2(t)}y^2 -2\frac{\big(\lambda-i\int B_3(t) dt \big)}{A_2(t)}y +i\frac{A_1(t)}{2\sqrt{A_2(t)}}x_1\right)\times\\
& \times\exp\left(\frac{ i}{2}\int \frac{ A_1(t)^4}{A_2(t)^4} dt-2 i\int\frac{ A_1(t)^2}{A_2(t)^3}G(x_2)dt\right)
\end{align*}

\item Reduced Lax pair
\begin{align}
&\Phi_{x_1x_1}-2(i\lambda -U_{x_1})\Phi=0\\
&\Phi_{x_2}- 4 (2i\lambda +U_{x_1})\Phi_{x_1}+2 U_{x_1x_1} \Phi =0
\end{align}
which is the Lax pair corresponding with the KdV equation.

Similarly, another constant of integration dubbed as ``$\lambda$'' arises. Again, this constant plays the role of the spectral parameter.

\end{itemize}

\item {\it Case II.2. $A_3(t)=0$, $A_1(t)\neq 0$, $A_2(t)=0$}
\begin{itemize}
\item Reduced variables
\begin{equation*}
x_1=y,\quad x_2=t.
\end{equation*}
\item Reduced fields
\begin{align*}
u(x,y,t)=&\frac{A_1(t)U(x_1,x_2)}{8 \left(\lambda-i\int B_3(t) dt \right)}+ \left(\frac{B_1(t)}{A_1(t)}+\frac{\dot A_1(t)}{4 A_1(t)}y \right)x + \frac{B_3(t)}{2 A_1(t)}x_1^2
\end{align*}
\begin{align*}
w(x,y,t)&=\frac{\Omega(x_1,x_2)}{2 i}+\frac{\dot A_1(t)}{8 A_1(t)}x^2+ \left (\frac{B_2(t)}{A_1(t)}+\frac{B_3(t)}{A_1(t)}y \right) x+\\
&+\frac{\dot A_1(t)^2+ A_1(t)\ddot A_1(t)}{24 A_1(t)^2}x_1^3+\frac{B_1(t)\dot A_1(t)+ A_1(t)\dot B_1(t)}{2 A_1(t)^2}x_1^2
\end{align*}
These reduced fields lead to a trivial reduction of the equations.

\item Reduced equation
\begin{align*}
&U_{x_1x_1}=0,\nonumber\\
&\Omega_{x_1x_1}=0
\end{align*}
\item Reduced eigenfunction
\begin{align*}
\Psi (x,y,t)=&\frac{\Phi (x_1,x_2)}{\sqrt{ A_1(t)}}\exp \{-\frac{2\left(\lambda-i\int B_3(t) dt \right)}{A_1(t)}x + i\frac{\dot A_1(t)}{4A_1(t)}x_1^2 + \\
+& 2i \frac{A_1(t)B_1(t)+2\left(\lambda-i\int B_3(t) dt \right)^2}{A_1(t)^2}x_1-\\
-&8i\int\frac{\left(B_1(t)A_1(t) +  2\left(\lambda -i\int B_3(t)dt\right)^2\right)^2}{A_1(t)^4}dt\}
\end{align*}
\item Reduced Lax pair
\begin{align}
&\Phi_{x_1}=0,\nonumber\\
&\Phi_{x_2}-\big(U_{x_1}+\Omega_{x_1}\big)\Phi =0
\end{align}
\end{itemize}

\item{\it Case II.3. $A_3(t)=0$, $A_1(t)=0$, $A_2(t)\neq 0$}

\begin{itemize}
\item Reduced variables
\begin{equation*}
x_1=\frac{x}{\sqrt{ A_2(t)}}-4\int \frac{B_1(t)}{ A_2(t)^{3/2}}dt\quad x_2=t,
\end{equation*}
\item Reduced fields
\begin{align*}
u(x,y,t)&=\frac{U(x_1,x_2)}{\sqrt{ A_2(t)}}+ \left(\frac{B_1(t)}{A_2(t)}+\frac{\dot A_2(t)}{8 A_2(t)}x \right) y+\frac{\int B_3(t) dt}{\sqrt{ A_2(t)}}x_1\\
w(x,y,t)&=\Omega(x_1,x_2)+\frac{\ddot A_2(t)}{48 A_2(t)}y^3+\frac{B_3(t)}{2 A_2(t)}y^2+\frac{B_2(t)}{A_2(t)}y
\end{align*}

These reduced fields lead to trivial a reduction of the equation.

\item Reduced equation
\begin{align}
U_{x_1x_2}=0
\end{align}
\item Reduced eigenfunction
\begin{align*}
\Psi (x,y,t)&=\frac{\Phi (x_1,x_2)}{ A_2(t)^{1/4}}\exp\left (i\frac{\dot A_2(t)}{8A_2(t)}y^2 -2\frac{\lambda-i\int B_3(t) dt}{A_2(t)}y\right)\nonumber\\
&\exp\left( 2i\int\frac{B_2(t)A_2(t) +  4\left(\lambda -i\int B_3(t)dt\right)^2}{A_2(t)^2}dt\right)
\end{align*}

\item Reduced Lax pair
\begin{align*}
&\Phi_{x_2}=0\\
&\Phi_{x_1x_1}+2 \big(U_{x_1}-i\lambda\big)\Phi =0
\end{align*}
\end{itemize}
\end{itemize}

%
%


\subsection{Reduction of a Lax pair in $1+1$ dimensions}

Let us now study the nontrivial reduction I.1. obtained in the past section.
We consider this reduction of interest from a possible physical viewpoint. In this way,
we aim to perform another symmetry search on the equation and its corresponding Lax pair in $1+1$ dimensions.

\medskip
We reconsider the equation given in \eqref{rode}
\begin{align}\label{cbkpr}
&U_{x_2x_2}=\Omega_{x_1x_2},\nonumber \\
&U_{x_1x_1x_1x_2}+8U_{x_1x_2}U_{x_1}+4U_{x_1x_1}U_{x_2}-\Omega_{x_2x_2}=0,
\end{align}
which is integrable in the Painlev\'e sense and possesses an associated linear spectral problem or Lax pair \eqref{rodelp}, which
takes the form
\begin{align}\label{plnontrred}
&\Phi_{x_1x_1}+i\Phi_{x_2}+2U_{x_1}\Phi=0,\nonumber \\
&i\Phi_{x_2x_2}-2U_{x_2}\Phi_{x_1}+\left(U_{x_1x_2}+i\Omega_{x_2}+\lambda\right)\Phi=0.
\end{align}
\noindent
whose compatibility condition $(\Phi_{x_1x_1x_2}=\Phi_{x_2x_2x_1})$ recovers \eqref{cbkpr}.
This Lax pair presents a constant parameter ``$\lambda$'' that plays the role of the spectral parameter. 

We aim at studying its classical Lie point symmetries and further reduction under the action of the symmetries.

\noindent
We propose a general transformation in which the infinitesimal generator depends on any independent and dependent variables as
\begin{align}\label{rbkpLietransf}
x_1&\rightarrow x_1+\epsilon \xi_1(x_1,x_2,U,\Omega)+O(\epsilon^2),\nonumber\\
x_2&\rightarrow x_2+\epsilon \xi_2(x_1,x_2,U,\Omega)+O(\epsilon^2),\nonumber \\
\lambda&\rightarrow \lambda+\epsilon \eta_{\lambda}(x_1,x_2,\lambda,\Phi)+O(\epsilon^2),\nonumber\\
U&\rightarrow U+\epsilon \eta_{U}(x_1,x_2,U,\Omega)+O(\epsilon^2),\nonumber\\
\Omega&\rightarrow \Omega+\epsilon \eta_{\Omega}(x_1,x_2,U,\Omega)+O(\epsilon^2).
\end{align}
Here we can see that we have considered {$\lambda$ as an independent variable} in order to make the reductions properly.
\noindent
If we want to achieve symmetries of equation \eqref{cbkpr} and those of the corresponding Lax pair \eqref{plnontrred} at the same time, we must include the transformed eigenfuncions
\begin{equation*}
\Phi \rightarrow \Phi+\epsilon \eta_{\Phi}(x_1,x_2,U,\Omega,\lambda,\Phi)+O(\epsilon^2),
\end{equation*}
where we have only specified the transformation for one of the eigenfunctions, since the complex conjugate version was not
considered in the former reductions. Similar results can be obtained for the complex conjugate by extension of previous results.

By definition of symmetry, transformation \eqref{rbkpLietransf} must leave invariant equation \eqref{cbkpr} and its associated Lax pair \eqref{plnontrred}.
The associated symmetry vector field has the expression
\begin{equation}\label{vfBKP11}
X=\xi_1\frac{\partial}{\partial x_1}+\xi_2\frac{\partial}{\partial x_2}+\eta_{\lambda}\frac{\partial}{\partial \lambda}+\eta_{U}\frac{\partial}{\partial U}+\eta_{\Omega}\frac{\partial}{\partial \Omega}+\eta_{\Phi}\frac{\partial}{\partial \Phi}.
\end{equation}
\noindent
From the terms in $\epsilon=0$ we retrieve the original equations,

\begin{align}\label{cbkpr2}
&U_{x_2x_2}=\Omega_{x_1x_2},\nonumber\\
&U_{x_1x_1x_1x_2}=\Omega_{x_2x_2}-8U_{x_1x_2}U_{x_1}-4U_{x_1x_1}U_{x_2},\nonumber\\
&\Phi_{x_1x_1}=-i\Phi_{x_2}-2U_{x_1}\Phi,\nonumber\\
&i\Phi_{x_2x_2}=2U_{x_2}\Phi_{x_1}-\left(U_{x_1x_2}-i\Omega_{x_2}-\lambda\right)\Phi
\end{align}

\noindent
that shall be used in the forthcoming steps.

First introduce transformation \eqref{rbkpLietransf} into the system of differential equations in \eqref{cbkpr}, \eqref{plnontrred} and
set the linear term in $\epsilon$ equal to zero. 
Introduce the needed prolongations for the equation \eqref{cbkpr}, that are $\eta_{U}^{[x_1x_1x_1x_2]}$, $\eta_{U}^{[x_1x_2]}$, $\eta_{U}^{[x_1x_1]}$, $\eta_{U}^{[x_2x_2]}$, $\eta_{U}^{[x_1]}$, $\eta_{U}^{[x_2]}$, $\eta_{\Omega}^{[x_2x_2]}$,
$\eta_{\Omega}^{[x_1x_2]}$, $\eta_{\Omega}^{[x_2]}$, and the prolongations neeeded for the Lax pair \eqref{plnontrred}, that are $\eta_{\Phi}^{[x_1x_1]}$, $\eta_{\Phi}^{[x_2x_2]}$, $\eta_{\Phi}^{[x_1]}$, $\eta_{\Phi}^{[x_2]}$, calculated following Lie's formula \cite{Stephani} and $U_{x_2x_2}$, $U_{x_1x_1x_1x_2}$, $\Phi_{x_1x_1}$, $\Phi_{x_2x_2}$ from \eqref{cbkpr2}.


%
%
We come up with the classical Lie point symmetries
\begin{align}\label{ncsymbkpr}
&\xi_1(x_1,x_2,U,\Omega)=\frac{1}{2}k_1x_1+k_2,\nonumber\\
&\xi_2(x_1,x_2,U,\Omega)=k_1x_2+k_3,\nonumber\\
&\eta_{\lambda}=-2k_1\lambda+k_4,\nonumber\\
&\eta_{u}(x_1,x_2,U,\Omega)=-\frac{1}{2}k_1U+k_5,\nonumber\\
&\eta_{\Omega}(x_1,x_2,U,\Omega)=-k_1\Omega+ik_4x_2+k_6(x_1),\nonumber\\
&\eta_{\Phi}(x_1,x_2,\lambda,U,W,\Phi,\lambda)=B(\lambda)\Phi.
\end{align}
These symmetries depend on 6 arbitrary constants of integration $k_1,k_2,k_3,k_4$, 
\newline
$k_5,C_0$ in $\mathbb{R}$ and two arbitrary functions $k_6(x_1)$
and $B(\lambda)$.

\begin{table}[H]\centering
\begin{tabular}{lllclll}\toprule
\multicolumn{3}{c}{Case  I: $k_1\neq 0$}&&\multicolumn{3}{c}{Case  II: $k_1=0$}
\\ 
\midrule
1.&$k_2\neq 0$& $k_3\neq 0$&&1.& $k_2\neq 0$&$k_3\neq 0$\\
& & &&2.& $k_2\neq 0$&$k_3=0$\\
& & &&3.& $k_2=0$&$k_3\neq 0$\\
\bottomrule
\end{tabular}
\caption{Reductions for $1+1$-BKP}
\label{Tab3}
\end{table}
\noindent
We introduce the following notation for the reduced variables. In this case, $\lambda$ is an independent variable.
\begin{equation}
x_1,x_2,\lambda \rightarrow z,\Lambda
\end{equation}
and the reduced fields and eigenfunctions

\begin{align}
U(x_1,x_2)&\rightarrow V(z),\nonumber\\
\Omega(x_1,x_2)&\rightarrow W(z),\nonumber\\
\Phi(x_1,x_2)&\rightarrow \varphi(z,\Lambda).
\end{align}


We find the following reductions

\begin{itemize}

\item{\it Case I.1. $k_1\neq 0$, $k_4\neq 0$}

\begin{itemize}

\item Reduced variables
\begin{align*}
&z={k_1 (k_1x_2+k_3) \over (k_1x_1+2 k_2)^2 },\\
&\Lambda=k_1^{-5} (2 k_1\lambda-k_4) (k_1x_1+2 k_2)^4.
\end{align*}

\item Reduced fields
\begin{equation*}
U(x_1, x_2)=k_1{ V(z) \over (k_1x_1+2 k_2)},
\end{equation*}
\begin{equation*}
\Omega(x_1, x_2)={ik_4\over 2 k_1^3}  (k_1x_1+2 k_2)^2 z-{k_1^2\over 2 (k_1x_1+2 k_2)^2} {W(z)\over z}.
\end{equation*}

\item Reduced Eigenfunction
\begin{equation*}
\Phi(x_1, x_2,\lambda)=\varphi (z,\Lambda)e^{\int{\frac{B(\lambda)}{k_4-2k_1\lambda}d\lambda}}
\end{equation*}

\item Reduced Lax pair
\begin{align*}
2 z^2 \varphi_{zz} +8 i z^2 V_z &( 2 \Lambda  \varphi_{\Lambda }- z \varphi_z)+\\
&+\Big [ W-z (W_z-6izV_z) +iz^2(4zV_{zz}-\Lambda) \Big] \varphi=0,
\end{align*}
\begin{align*}
16 \Lambda^2  \varphi_{\Lambda\Lambda}&-16 z \Lambda   \varphi_{z\Lambda}+4 \Lambda(3-8 i z^2 V_z)\varphi_{\Lambda }+(i+6 z +16 i z^3 V_z) \varphi_z- \\
&-2 \Big[W+ V+ z (1+6iz) V_z+iz^2(4zV_{zz}- \Lambda) \Big]\varphi=0.
\end{align*}

In this case, a constant ``$\Lambda$'' appears and plays the role of a spectral parameter.

\item Reduced equations

\begin{equation*}
W_{zz}=V_{zz},
\end{equation*}
\begin{align*}
16z^6V_{zzzz}+&144z^5V_{zzz}+8z^3(15-8V)V_z+2zW_z-2 W-\\
&\hspace{-0.5em}-(1-300z^2+32z^2V+96z^3V_z)z^2V_{zz}-176z^4V_z^2=0.
\end{align*}

\end{itemize}

\item{\it Case II.1. $k_1=0$, $k_2\neq 0$, $k_3\neq 0$}

\begin{itemize}

\item Reduced variables
\begin{align*}
&z=\frac{k_2}{k_3}\left(\frac{k_2}{k_3}x_2-x_1\right)\\
&\Lambda=\frac{k_2}{k_3}\left(\frac{k_2}{k_4}\lambda-x_1\right).
\end{align*}
\item Reduced fields
\begin{align*}
&U(x_1,x_2)=\frac{k_2}{k_3}V(z)+\frac{k_5}{k_2}x_1,\\
&\Omega(x_1,x_2)=-\frac{k_2^2}{k_3^2}W(z)+i\frac{k_4k_3^2}{k_2^3}\left(z+\frac{k_2}{2k_3}x_1\right)x_1+\frac{1}{k_2}\int{k_6(x_1)dx_1}.
\end{align*}

\item Reduced Eigenfunction
\begin{equation*}
\Phi(x_1,x_2,\lambda)=\varphi(z,\Lambda)e^{\frac{1}{k_4}\int{B(\lambda)d\lambda}}.
\end{equation*}

\item Reduced Lax pair

\begin{align}
&\varphi_{\Lambda \Lambda}+2\varphi_{z \Lambda}+\varphi_{zz}+i\varphi_z+2\left(C_1-V_z\right)\varphi=0,\nonumber\\
&(-iC_2\Lambda-W_z+iV_{zz})\varphi-2i\left(\varphi_{\Lambda}+\varphi_z\right)V_z+\varphi_{zz}=0.
\end{align}

In this reduced spectral problem, the constant ``$\Lambda$'' plays the role of the spectral parameter.
\item Reduced equation

\begin{align}
&W_{zz}=V_{zz}-iC_1,\nonumber\\
&V_{zzzz}+4\left(2C_2-3V_z\right)V_{zz}-W_{zz}=0.
\end{align}

where $C_1=\frac{k_3^2k_5}{k_2^3}$ and $C_2=\frac{k_3^5k_4}{k_2^6}$
\end{itemize}

\item{\it Case II.2. $k_1=0$, $k_2\neq 0$, $k_3=0$, $k_4\neq 0$, $k_5\neq 0$}
\begin{itemize}
\item Reduced variables
\begin{align*}
&z=\frac{k_5}{k_2}x_2,\nonumber\\
&\Lambda=\left(\frac{k_5}{k_2}\right)^{1/2}\left(\frac{k_2}{k_4}\lambda-x_1\right)
\end{align*}
\item Reduced fields
\begin{align*}
&U(x_1,x_2)=\left(\frac{k_5}{k_2}\right)^{1/2}V(z)+\frac{k_5}{k_2}x_1,\nonumber\\
&\Omega(x_1,x_2)=\frac{k_5}{k_2}W(z)+i\frac{k_4}{k_2}x_2x_1+\frac{1}{k_2}\int{k_6(x_1)dx_1}.
\end{align*}
\item Reduced Eigenfunction
\begin{equation*}
\psi(x_1,x_2)=\varphi(z,\Lambda)e^{\frac{1}{k_4}\int{B(\lambda)d\lambda}}.
\end{equation*}
\item Reduced Lax pair

\begin{equation*}
\varphi_{\Lambda \Lambda}+i\varphi_{z}+2\varphi=0,
\end{equation*}
\begin{equation}
\varphi_{zz}-2iV_z\varphi_{\Lambda}+\left(W_z-\Lambda V_{zz}\right)\varphi=0.
\end{equation}

\item Reduced equation

\begin{align}
       &V_{zz}=i\frac{k_2^{3/2}k_4}{k_5^{5/2}},\nonumber\\
       &W_{zz}=0.
       \end{align}

\end{itemize}


\item{\it Case II.3. $k_1=0$, $k_2=0$, $k_3\neq 0$, $k_4\neq 0$, $k_5\neq 0$}
\begin{itemize}
\item Reduced variables
\begin{align*}
&z=\left(\frac{k_5}{k_3}\right)^{1/3}x_1,\nonumber\\
&\Lambda=\frac{k_3^{1/3}k_5^{2/3}}{k_4}\left(\lambda-\frac{k_4}{k_3}x_2\right)+i\left(\frac{k_5}{k_3}\right)^{2/3}\frac{B_0}{k_4}
\end{align*}
\item Reduced fields
\begin{align*}
&U(x_1,x_2)=\left(\frac{k_5}{2k_3}\right)^{1/3}V(z)+\frac{k_5}{k_3}x_2,\nonumber\\
&\Omega(x_1,x_2)=W(z)+i\frac{k_4}{2k_3}x_2^2+\frac{B_0}{k_3}x_2.
\end{align*}
where $B_0$ is a constant.
\item Reduced Eigenfunction
\begin{equation*}
\psi(x_1,x_2)=\varphi(z,\Lambda)e^{\frac{1}{k_4}\int{B(\lambda)d\lambda}}.
\end{equation*}
\item Reduced Lax pair

\begin{equation*}
\varphi_{\Lambda \Lambda}+2i\varphi_{z}-2\Lambda V_{zz}\varphi=0,
\end{equation*}
\begin{equation*}
\varphi_{zz}-i\varphi_{\Lambda}+V_z\varphi=0.
\end{equation*}
Here, the constant ``$\Lambda$'' plays the role of the spectral paramenter.
\item Reduced equation

\begin{align}
       &V_{zz}=i\frac{k_3k_4}{2k_5^2},\nonumber\\
       &W_{zz}=0.
       \end{align}

\end{itemize}

\end{itemize}

\section{Conclusions and brief comments}

We have calculated the classical (point) symmetries of a nonlinear PDE and its corresponding two-component, nonisospectral Lax pair in $2+1$ dimensions. 
The spectral parameter has been conveniently introduced as dependent variable obeying the nonisospectral condition.
We have reduced both the equation and its Lax pair attending to different choices of the arbitrary functions 
and constants of integration present in the calculated symmetries. 
Of a total of 6 reductions, 2 of them happen to be knowledgeable equations in the literature. 
One is the KdV equation. {From this result, we can say that ($2+1$)-BKP is a generalization of the KdV.}
The nonclassical version when $\xi_3\neq 0$ has also been obtained, leading to the same set of symmetries (although such results have not been included in this manuscript).
The other interesting reduction is an equation showing multisoliton solutions.
In this equation and its associated Lax pair, we have performed a second classical Lie symmetry analysis.
In this case, the spectral parameter has been conveniently introduced in the reduction and it needs to be considered as an independent variable.
Another 4 possible reductions are contemplated.

After the computation of the classical Lie symmetry results, the nonclassical symmetries for the nontrivial reduction I.1. studied in Section 3.1. \eqref{cbkpr} and those
of its corresponding Lax pair \eqref{plnontrred} have also
been obtained, provided that $\xi_2\neq 0$, leading us to equivalent results (not listed in this paper). {Therefore, the classical and nonclassical symmetries of equations  \eqref{cbkpr} and \eqref{plnontrred} are equivalent.
Nonetheless, this is not always the rule. Indeed, nonclassical symmetries are usually more general and contain the classical ones as a particular case. }

As a future perspective, it would be desirable to be able to find ways of solving nonisospectral Lax pairs in $1+1$ dimensions, since the
methods applicable are only applicable in the cases of isospectral ones.

\section*{Acknowledgements}
P.G. Est\'evez, J.D. Lejarreta and C. Sard\'on acknowledge partial financial support by research projects
MAT2013-46308-C2-1-R (MINECO) and SA226013 (JCYL).

\end{document}